\documentclass[12pt]{article}
\pdfoutput=1
\usepackage{jheppub}
\usepackage{amsmath}
\usepackage{amssymb}
\usepackage{graphicx}
\usepackage{float}
\usepackage{mathtools}
\usepackage{caption}
\usepackage{amsthm}
\usepackage{wrapfig}
\usepackage{xcolor}
\usepackage{slashed}
\usepackage{hyperref}
\usepackage{tikz}
    \usetikzlibrary{angles, calc, cd}
    \usetikzlibrary{arrows.meta, tikzmark}

\newcommand{\cn}{{\cal N}}
\def\bal#1\eal{\begin{align}#1\end{align}}

\title{$\cn = (0, 2)$ higher-spin supergravity in AdS$_3$}

\author[1]{Zisong Cao}

\affiliation[1]{ Kavli Institute for Theoretical Sciences (KITS)\\
University of the Chinese Academy of
Sciences, Beijing 100190, China}

\emailAdd{caozisong20@mails.ucas.ac.cn}

\begin{document}

\begin{minipage}{\textwidth}
\flushright{USTC-ICTS/PCFT-25-}
\end{minipage}

\vspace{-2.8em}

\abstract{In this paper we discuss the higher-spin gravity in 3d with supersymmetries $\cn = (0, 2)$, by which we mean that the asymptotic symmetry of such a gravity theory have the structure of 2d $\cn = (0, 2)$ superconformal algebra.
We generalized the Vasiliev equation in 3d that appear as subsector in theories with higher-spin symmetry. Asymptotic symmetry and possible matter content of such generalized theories is discussed.
Also, the 1-loop partition function of this theory around thermal Euclidean AdS space-time, with different matter fields, is calculated by heat-kernel method. All the construction and discussion is limited to linearized level and apply to full theories with corresponding higher-spin gauge symmetries.}

\maketitle

\section{Introduction}
Higher-spin gravity theory in AdS space-time has long been studied as it describes interactions between massless spin-$s\geq 2$ fields, bypassing wide-known no-go theorems~\cite{Coleman:1967ad, Bekaert:2010hw,Maldacena:2011jn,Maldacena:2012sf}
against such theories.
Since gravity should be contained in higher-spin gravity theories, they are expected to be dual to some CFT with corresponding higher-spin currents by the AdS/CFT conjecture~\cite{Maldacena:1997re}. Therefore, higher-spin gravity theories are often regarded as toy models of holographic duality.
Many works on holography of higher-spin gravity have already been done in different dimensions and with different amount of supersymmetry~\cite{Klebanov:2002ja, Gaberdiel:2010pz, Gaberdiel:2012uj, Creutzig:2011fe, Candu:2012jq, Creutzig:2014ula}.
Abstractly, for gravity theories with dimension equal to or greater than $3+1\mathrm{d}$, a series of theorems that in analog with Coleman-Weinberg theorem~\cite{Maldacena:2011jn,Maldacena:2012sf,Boulanger:2013zza,Alba:2013yda,Alba:2015upa}, severely
constraints the structure of holographic AdS higher-spin gravity theories hardly, forcing them to be dual with (singlet sector of) free theories.
In the other side, for $3\mathrm{d}/2\mathrm{d}$ holography, where these Coleman-Weinberg-like theorems fall down, there are much more freedom to construct theories.

Classically, as higher-spin gauge symmetry is given, the simplest gravity theory with matter is the so-called Vasiliev higher-spin gravity theory, where the action of higher-spin fields are given by the Chern-Simons formalism, while matter fields are added in by
a tower of 1-order differential equations.
In practice, Vasiliev higher-spin gravity theories in $2+1\mathrm{d}$ are conjectured to be dual to the `t Hooft limit of 2d minimal model CFTs up to 1-loop order~\cite{Gaberdiel:2010pz, Gaberdiel:2012uj, Creutzig:2011fe, Candu:2012jq}, which are interaction CFTs with higher-spin currents.
As conserved currents, including the (super)conformal currents, in $2\mathrm{d}$ CFTs are either
``holomorphic'' (left-moving) or ``anti-holomorphic'' (right-moving),
theories with different numbers of supersymmetries, e.g. $\cn = (p, q)$, on the two sides can be constructed~\cite{Melnikov:2019tpl}. Also, such $\cn = (p, q)$ (super)conformal symmetry can be realized as asymptotic symmetry of certain supergravity theories in AdS$_3$~\cite{Achucarro:1986uwr, Achucarro:1989gm}.
Recently, higher-spin symmetry with $\cn = (0, 2)$ supersymmetry was observed in the IR fixed point of some double-scaled 2d disordered models~\cite{Peng:2018zap}, see~\cite{Chang:2021fmd,Chang:2021wbx} for similar observations in 3d.
Therefore, while most previous work on $3\mathrm{d} \text{ HS}/2\mathrm{d} \text{ CFT}$ holography focus on theories with equal amount of supersymmetry on left-moving and right-moving sectors~\cite{Creutzig:2011fe,Candu:2012jq,Henneaux:2012ny,Hanaki:2012yf,Datta:2012km,Peng:2012ae,Gaberdiel:2013vva,Gaberdiel:2014yla}, 
it is meaningful to look for a higher-spin supergravity in AdS$_3$ that possibly possessing $\cn = (0, 2)$ supersymmetry. In this paper we construct explicitly such higher-spin theories with $\cn = (0, 2)$ supersymmetry to the linearized level.

Subtleties about the Vasiliev higher-spin theory at the quantum level should be clarified here. There are some problems when one tries to quantize the Vasiliev theory itself, especially when going beyond the linear order.\footnote{We thank Prof. Skvortsov to point out this.}
It is argued that whenever corrections beyond linear order are considered, additional fields that have not been understood well in the AdS/CFT aspect will appear~\cite{Kessel:2015kna, Sharapov:2024euk}. Therefore, Vasiliev theory should be regarded as a subsector of some more general higher-spin gravity theory.
For example, it is believed that Vasiliev theory in $3\mathrm{d}$
appears as a subsector of string theory in tensionless limit~\cite{Gaberdiel:2013vva}, where its higher-spin gauge symmetry becomes the first Regge trajectory of the enhanced higher spin square (HSS) symmetry~\cite{Gaberdiel:2014cha}. Also, by taking ``twisted'' fields into consideration, a construction via Poisson sigma model is suggested recently~\cite{Sharapov:2024euk}.
In our paper, we use the Vasiliev theory to the linearized level, by which the discussion relies only on the representation theory and AdS geometry, therefore our result applies to any full theory containing the Vasiliev theory as a subsector.

A brief overview of this paper is as follows: In Section.\ref{sec:review} the construction of Vasiliev higher-spin gravity in $2\mathrm{d}$ is reviewed.
Then in Section.\ref{sec:construction} we construct the desired (linearized) $\cn = (0, 2)$ higher-spin gravity theory, discuss its asymptotic symmetry and possible matter fields content. In Section.\ref{sec:one_loop_partfunc} we derive the formular of 1-loop partition functions around thermal AdS space-time with different choice of matter contents, containing a heat-kernel calculation of contribution from a kind of fermions that have not been calculated before.
Finally in Section.\ref{sec:discussion} we give a brief conclusion to the result of this paper and look into further possible researches.

\section{Vasiliev higher-spin supergravity}\label{sec:review}
In this section we review the construction for linear version of Vasiliev higher-spin gravity theory with $\cn=2$ supersymmetry~\cite{Vasiliev:1999ba}, based on the general formalism developed in e.g.~\cite{Vasiliev:1992gr, Prokushkin:1998bq, Prokushkin:1998vn, Vasiliev:1999ba, Barabanshchikov:1996mc}.

\subsection{Higher-spin theory }

Let us first recall briefly some fundamental concepts of higher-spin gravity theory in $3\mathrm{d}$.
Massless higher-spin fields with spin-$n$ greater than two can be described~\cite{Fang:1978wz, Fronsdal:1978rb} as gauge fields with  gauge transformation
\begin{equation}
    \delta\Psi_{\mu_1\dots\mu_n} = \partial_{(\mu_1}\xi_{\mu_2\dots\mu_n)}\,,
\end{equation}
where $\xi$ is traceless and $\Psi$ is symmetric and double-traceless ${\Psi_{\mu_1\dots\mu_{n-4}\lambda\rho}}^{\lambda\rho}$,
and the gauge connection appears in the covariant derivative $D_{\mu}$. $\Psi$ is sometimes refered to as metric-like fields, and the metric $g_{\mu\nu}$ can be regarded as the $n=2$ case of it. 
Beside the metric-like form, higher-spin fields in $3\mathrm{d}$ can also be presented in the ``frame-like'' form.
For example, the metric $g_{\mu\nu}$ is encoded in the frame field $e_{\mu}^a$ and local Lorentz connection $\omega_\mu^{a} = \frac{1}{2}\epsilon^a_{\; bc}\omega_{\mu}^{bc}$, with $a,b,c$ local Lorentz indices. 
Generally, higher-spin fields of spin-$(n+1)$ are presented as the higher-spin generalization of the two gauge fields $e_{\mu}^a$ and $\omega_\mu^{a}$~\cite{Campoleoni:2010zq}
\begin{equation}
    {e_{\mu}}^{a_1\dots a_{n}}\,, \qquad  {\omega_{\mu}}^{a_1\dots a_{n}}\,,
\end{equation}
whose gauge transformations are
\bal
    \delta {e_{\mu}}^{a_1\dots a_{n}} &= D_\mu \xi^{a_1\dots a_n} + \epsilon^{bc(a_1|}e_{\mu, b}{\Lambda_{c}}^{|a_2\dots a_n)}\\
    \delta {\omega_{\mu}}^{a_1\dots a_{n}} &= D_\mu \Lambda^{a_1\dots a_n} + \frac{1}{l_{\text{AdS}}^2}\epsilon^{bc(a_1|}e_{\mu, b}{\xi_{c}}^{|a_2\dots a_n)}\ .
\eal

One can combine $e_\mu$ and $\omega_\mu$ into
\begin{equation}
    A = \omega + \frac{1}{l_{\mathrm{AdS}}} e\,, \quad \bar{A} = \omega - \frac{1}{l_{\mathrm{AdS}}} e\,,
\end{equation}for a simpler form of gauge transformation,
where $l_{\mathrm{AdS}}$ is the AdS radius related to the cosmological constant by $\lambda = -\frac{1}{l_{\mathrm{AdS}}^2}$.
Now $A$ and $\bar{A}$ can be regarded as some Lie algebra valued 1-form, with which one may code interactions of higher-spin gauge fields into gauge theories.
Additionally, since in $3\mathrm{d}$ higher-spin theory has no propagating degrees of freedom, one expect the gauge theory to be a topological theory.
In fact, it has been known for long that the action of 3d pure gravity (and supergravity) with a negative cosmological constant can be formulated as a Chern-Simons theory~\cite{Witten:1988hc, Achucarro:1986uwr, Achucarro:1989gm,Witten:2007kt}.
Similar formulation has been established for various higher-spin theory in 3d~\cite{Bekaert:2004qos, Gaberdiel:2010ar,Campoleoni:2010zq, Castro:2010ce,Gaberdiel:2011wb,Gutperle:2011kf,Gaberdiel:2011zw,Ammon:2011nk,Ammon:2011ua,Creutzig:2011fe, Tan:2012xi}, see also~\cite{Chang:2011mz,Jevicki:2011ss}.
Generally speaking, each Lie algebra $\mathfrak{g} = \mathfrak{g}_{\text{L}}\oplus \mathfrak{g}_{\text{R}}$ with trace and an injection from $\mathrm{sl}(2, \mathbb{R}) \oplus \mathrm{sl}(2, \mathbb{R})$ 
defines an interacting higher-spin theory.
For example, one can construct the $\mathrm{SL}(N)\times\mathrm{SL}(N)$ higher-spin theory with higher-spin fields up to spin $s=N$~\cite{Campoleoni:2010zq}.
However, consistency with the chaos bound~\cite{Maldacena:2015waa} seems to require an infinite tower of higher-spin fields, at least in the weakly interacting region~\cite{Perlmutter:2016pkf, Narayan:2019ove}.

In addition to higher-spin gauge fields, we also include
the matter sector in higher-spin gravity theories, which furnishes nontrivial representations of the higher-spin gauge symmetry.
It is convenient to describe the matter fields in the ``unfolded formalism'', where matter fields of these higher-spin gravity theory satisfies the `unfolded' Klein-Gordon equation (for spinors, Dirac equation)~\cite{Barabanshchikov:1996mc, Prokushkin:1998bq}
that are series of first-order differential equations. These equations reduce to the traditional Klein-Gordon equation (Dirac equation) of scalars (spinors) once  all higher-spin components are eliminated while keeping the lowest-spin part.

In the following we construct the matter sector by modifying the Vasiliev equations with $\cn = 0$ or $\cn = 2$ supersymmetry~\cite{Vasiliev:1999ba, Prokushkin:1998bq}.
As is mentioned above, Vasiliev theories with such supersymmetries are expected to be dual with `t Hooft limits of $2\mathrm{d}$ minimal models~\cite{Gaberdiel:2012uj, Candu:2012jq}.
Though in general there are only linearized description of Vasiliev theories so far as we know, it is enough if one care about only 1-loop calculations around some backgrounds.
Start with the linearized description of $\cn = 2$ Vasiliev theory and the dual description by 2$\mathrm{d}$ minimal model CFT, we project out fermionic higher-spin fields that are the super-descendants of the bosonic degrees of freedom  in the left-moving sector and then discuss representation theory of the remained gauge Lie algebra, deriving the possibility of matter supermultiplets of this gauge algebra. We find that the gauge Lie algebra have the structure of 
\[
{\mathrm{shs}[(1-\nu)/2]}_{\text{L}} \oplus {\left(\mathrm{hs}[(1-\nu)/2]\oplus \mathrm{hs}[(1+\nu)/2]\oplus \mathrm{u}(1)\right)}_{\text{R}}
\]with $\nu$ some parameter, and that possible matter fields falls into four choice of supermultiplets, where mass of components is determined by the $\nu$ parameter. Therefore spectrum of such theories is known.
In addition, 1-loop partition function of the theory around thermal $\text{EAdS}_3$ vacuum, with different choice of matter components, are calculated
by the heat-kernel method, with a bit of generalization of results in~\cite{Giombi:2008vd, David:2009xg,Creutzig:2011fe}. 
Among these analysis only linearized theory around AdS vacuum are used, showing the generality of these results for different possible non-linear completions.

In the remain part of this paper, we use the frame field $e_{\mu}^{\;a}$ to convert between space-time indices and local Lorentz indices, and spin-connections of local Lorentz symmetry $\omega_{\mu}^{\;a} := \frac{1}{2}\omega_{\mu}^{\;bc}\epsilon^a_{\;bc}$ to construct covariant derivatives $D_{\mu}$.
We further replace local Lorentz indices by spin indices $\alpha\beta\gamma\dots, \alpha\dots\in \{1, 2\}$ with gamma matrices $\gamma_a^{\;\alpha\beta}$:
\begin{equation}
     T_{\alpha_1\alpha_2\alpha_3\alpha_4\dots} = \gamma^a_{\;\alpha_1\alpha_2} \gamma^b_{\;\alpha_3\alpha_4} \cdots T_{a,b,\dots},
\end{equation}so that each spin-$s$ tensor have symmetric $2s$ spin indices. Therefore, we can deal with both bosonic and fermionic higher-spin fields in the same time.

\subsection{Unfolded formalism: higher-spin theory with matter fields}
We first describe the linear theory around certain vacuum. The action of higher-spin fields can be constructed by regarding them as gauge fields and using the Chern-Simons formalism. 
What we need is a Lie algebra $\mathfrak{g} = \mathfrak{g}_{\mathrm{L}} \oplus \mathfrak{g}_{\mathrm{R}}$
containing subalgebra $\mathrm{so}(2, 2) \cong \mathrm{sl}(2, \mathbb{R}) \oplus \mathrm{sl}(2, \mathbb{R})$, 
where both factor of $\mathfrak{g}$ contains one $\mathrm{sl}(2, \mathbb{R})$, and a trace on it. 
\subsubsection{Gauge fields}
In Vasiliev higher-spin gravity theory, an infinite dimensional associative algebra $\mathrm{Aq}(2,\nu)$ is used to generate Lie algebra structure and its representation. 
Concretely, $\mathrm{Aq}(2,\nu)$ is generated by deformed oscillators$y_\alpha$ and idempotent element $K$~\cite{Vasiliev:1999ba}:
\begin{equation}
    [y_\alpha, y_\beta] = 2\mathrm{i}\epsilon_{\alpha\beta}(1 + \nu K), \{K, y_{\alpha}\} = 0, K^2 = 1
\end{equation}
with a $\mathbb{Z}_2$ grading $\pi_f$ serving as the fermion parity:
\begin{equation}
    \pi_f(y_\alpha) = 1, \pi_f(K) = 0. 
\end{equation}
In general, elements of this algebra are represented as:
\begin{equation}\label{eq:gauge_field_component}
    f(y,K) = \sum_{n=0}^{\infty}\sum_{I,J=0,1} \frac{1}{n!} f^{\alpha_1\dots\alpha_n}_{I}{(K)}^I y_{(\alpha_1}\dots y_{\alpha_n)}.
\end{equation}This associative algebra allows a natural invariant bilinear form $\mathrm{str}$:
\begin{equation}
    \mathrm{str}(f(y,K)) = f_{I=0} - \nu f_{I=1}.
\end{equation}Especially, $\mathrm{str}(1) = 1, \mathrm{str}(K) = -\nu$. Non-zero spin part of $f(y, K)$ gives no contribution to $\mathrm{str}f(y, K)$.
In addition, traceless part of $Aq(2, \nu)$ is closed with Lie bracket, hence determines a Lie algebra.
In order to ensure the Lie algebra contains $\mathrm{so}(2, 2) \simeq \mathrm{sl}(2, \mathbb{R}) \oplus \mathrm{sl}(2, \mathbb{R})$ as a subalgebra, an additional central element $\psi$ is introduced such that:
\begin{equation}
    \pi_f(\psi) = 0, \psi^2 = 1, [\psi, K] = [\psi, y_{\alpha}] = 0.
\end{equation}
then,
\begin{equation}\label{eq:poincare_algebra}
    L_{\alpha\beta} = \frac{1}{4\mathrm{i}} \{y_{\alpha}, y_{\beta}\}; P_{\alpha\beta} = \frac{1}{4\mathrm{i}} \{y_{\alpha}, y_{\beta}\} \psi,
\end{equation}specifying the $\mathrm{so}(2, 2) \simeq \mathrm{sl}(2, \mathbb{R}) \oplus \mathrm{sl}(2, \mathbb{R})$ subalgebra.
Besides, $K^2 = \psi^2 = 1$ allow us to define two projection operator:
\begin{eqnarray}
    P^K_{\pm} = \frac{1}{2}(1\pm K); P^\psi_{\pm} = \frac{1}{2}(1 \pm \psi),
\end{eqnarray}dividing $f(y,K,\psi)$ into four part. Here $P^\psi_{\pm}$ serving as a ``chiral'' operator, specifying the ``left-moving'' $\psi = +$ and ``right-moving'' $\psi = -$ sector of fields
\footnote{While in 2+1d space-time chirality do not make sense, by holography, gauge fields in these two sectors corresponds to left and right moving sector of 2d CFT. 
That's way sometimes we may call themselves also ``left (right)-moving'' sector, though in $2+1\mathrm{d}$.}, 
while $P^K_{\pm}$ project out the n-odd or n-even elements of $f(y,K,\psi)$.

Now 1-form field $W(y,K,\psi|x)$ valued in the extended $\mathrm{Aq}(2, \nu) \otimes \langle 1, \psi \rangle$ with supertrace $\mathrm{str}(W) = 0$ defines the higher-spin sector of Vasiliev higher-spin gravity. Explicitly, these higher-spin and gravity fields are encoded in $W(y,K,\psi|x)$ as :
\begin{equation}
    W(y,K,\psi|x) = \sum_{J=0,1}W(x)\left(\nu + K\right){(\psi)}^J + \sum_{n=1}^{\infty}\sum_{I,J=0,1} \frac{1}{n!} W^{\alpha_1\dots\alpha_n}_{I}{(K)}^I y_{(\alpha_1}\dots y_{\alpha_n)}{(\psi)}^J.
\end{equation}
Especially, the $I=0$ and $n=2$ components of $W(y,K,\psi|x) $ represents to the gravity field. With Poincare algebra recognized as Eq.~\eqref{eq:poincare_algebra}, it is
\begin{equation}
    W^{\text{gr}}_\mu(y, K, \psi|x) = \frac{1}{8\mathrm{i}}(\omega_\mu^{\alpha\beta} + \lambda e^{\alpha\beta}\psi )y_{\alpha}y_{\beta},
\end{equation}
where gravity is represented by the frame field and local Lorentz connection that serves as gauge connection of $L_{\alpha\beta}$ and $P_{\alpha\beta}$:\footnote{Note that the definition of $P_{\alpha\beta}$ here is different with \cite{Witten:1988hc} by a $1/\sqrt{\lambda}$ factor. Therefore the corresponding gauge connection field should be not $\sqrt{\lambda} e$ as in \cite{Witten:1988hc} but $\lambda e$.}
\begin{equation}
    e_\mu^{\alpha\beta} = {(\gamma_a)}^{\alpha\beta}e_\mu^a,\qquad 
    \omega_\mu^{\alpha\beta} = \frac{1}{2} {(\gamma_a)}^{\alpha\beta} \epsilon^a_{\;\; b c} \omega_\mu^{\;\;b c}.
\end{equation}

Structure of $\cn = 2$ supersymmetry is given by ~\cite{Vasiliev:1999ba}:
\begin{equation}
    Q^{(1)}_{\alpha} = y_{\alpha}; Q^{(2)}_{\alpha} = y_{\alpha}K.
\end{equation}

The associative algebra $\mathrm{Aq}(2, \nu)$ can also be formulated as a quotient of $U(\mathrm{osp}(2,1))$, the universal envelope algebra of $\mathrm{osp}(2,1)$, as:
\[
\mathrm{Aq}(2, \nu) \cong U(\mathrm{osp(2,1)})/I_{(C_2-\nu^2)},
\]
with $C_2$ representing the Casimir element of $\mathrm{osp(2,1)}$,
clarifying the equivalence between the Lie algebra constructed above and the $\mathrm{shs}[\mu]$ algebra~\cite{Gaberdiel:2012uj, Candu:2012jq} with parameter$\mu = (\nu - 1)/2$:
\begin{equation}
    \mathrm{Aq}(2, \nu) \cong \mathbb{C} \oplus \mathrm{shs}[(1-\nu)/2].
\end{equation}

After representation decomposition,
\begin{equation}
    \mathrm{Aq}(2, \nu) = \mathbf{1} \oplus \sum_{j=0}^\infty(V^{(j)}\oplus V^{(j+1/2)}\oplus V^{(j+1/2)}\oplus V^{(j+1)}),
\end{equation}
giving two infinite tower of higher-spin $\cn = 2$ supermultiplet, containing fields with spin $(s, s+1/2, s+1/2, s+1) $ where $s=1, 2, 3, \dots$.
\subsubsection{Matter fields}
Beside higher-spin fields, a higher-spin gravity theory should also contain certain matter fields. In linearized gravity theory, as higher-spin (and gravity) fields being treated as backgrounds, matter fields are just required to be representations of their gauge algebra. It restrict the possibility of matter contents into several choices.
For matter fields in higher-spin gravity theory, the ``unfolded'' formalism is needed to write down their E.o.M, where every matter field should be accompanied by auxiliary fields with an infinite tower of indices, being similar to the gauge field Eq.~\eqref{eq:gauge_field_component}, where the lowest component refers to the original matter field, and satisfying some integrability conditions that giving desired E.o.M of original matter fields when restricted to the lowest component.
For example, matter field in unfolded formalism can be given by a function values in $\mathrm{Aq}(2, \nu)$ algebra:
\begin{equation}
    C(y, K |x) = \sum_{n=0}^\infty\sum_{I,J=0,1}\frac{1}{n!}C_{IJ}^{\alpha_1\dots\alpha_n}(x){(K)}^I y_{(\alpha_1}\dots y_{\alpha_n)}.
\end{equation}

Since the gauge Lie algebra of Vasiliev higher-spin gravity is constructed from the associative algebra $\mathrm{Aq}(2, \nu)$, there are automatically two representations of this Lie algebra by acting it on $\mathrm{Aq}(2, \nu)$ from left (right), called as ``left (right)'' canonical representation. 
Besides, the direct plus structure respected to $1\pm\psi$ doubles the number of choices. 
Therefore, representations of matter fields are built by the left (right) canonical representation of the left (right) part of lie algebra.
However, only those representations that decomposes into finite dimension representations of Lorentz symmetry $\mathrm{sl}(2, \mathbb{R})$ make sense for our application. 
Therefore left and right canonical representations should be tensor producted together into four combinations, corresponding to four projection conditions and gauge transformations for the $Aq(2, \nu)$ valued ``unfolded'' matter field:
\begin{equation}
    C(y, K, \psi|x) = \sum_{n=0}^\infty\sum_{I,J=0,1}\frac{1}{n!}C_{IJ}^{\alpha_1\dots\alpha_n}(x){(K)}^I y_{(\alpha_1}\dots y_{\alpha_n)}{(\psi)}^J,
\end{equation}
\begin{itemize}
    \item left canonical representation of left-moving sector direct plus right canonical representation of right-moving sector, represented as
    \begin{equation}
        \begin{aligned}
        P^\psi_{-}C(y,K,\psi|x) =&\; 0,  \\
        D C(y, K, \psi| x) =&\; d C(y, K, \psi|x)\\
        &\; + W(y, K, \psi| x)\wedge C(y, K, \psi| x) - C(y, K, \psi| x)\wedge W(y, K, -\psi| x);
        \end{aligned}
    \end{equation}
    \item left canonical representation of right-moving sector direct plus right canonical representation of left-moving sector, represented as
    \begin{equation}
        \begin{aligned}
        P^\psi_{+}C(y,K,\psi|x) =&\; 0,\\
        D C(y, K, \psi| x) =&\; d C(y, K, \psi|x)  \\
        &\; + W(y, K, \psi| x)\wedge C(y, K, \psi| x) - C(y, K, \psi| x)\wedge W(y, K, -\psi| x);
        \end{aligned}
    \end{equation}
    \item left canonical representation of left-moving sector direct plus right canonical representation of left-moving sector, represented as
    \begin{equation}
        \begin{aligned}
        P^\psi_{-}C(y,K,\psi|x) =&\; 0,  \\
        D C(y, K, \psi| x) =& \; d C(y, K, \psi|x) \\
        &\; + W(y, K, \psi| x)\wedge C(y, K, \psi| x) - C(y, K, \psi| x)\wedge W(y, K, \psi| x);
        \end{aligned}
    \end{equation}
    \item left canonical representation of right-moving sector direct plus right canonical representation of right-moving sector, represented as
    \begin{equation}
        \begin{aligned}
        P^\psi_{+}C(y,K,\psi|x) =&\; 0,  \\
        D C(y, K, \psi| x) =& \; d C(y, K, \psi|x)\\
        &\; + W(y, K, \psi| x)\wedge C(y, K, \psi| x) - C(y, K, \psi| x)\wedge W(y, K, \psi| x),
        \end{aligned}
    \end{equation}
\end{itemize}where $D$ represents covariant derivative, determining the integrability condition satisfied by unfolded matter fields.
By introducing another central element $\psi_2$ and denote the original $\psi$ as $\psi_1$, that:
\begin{equation}
    \pi_f(\psi_2) = 0, \psi_2^2 = 1; [\psi_2, y_\alpha] = [\psi_2, K] = 0; \{\psi_1, \psi_2\} = 0,
\end{equation}we can combine the four representations together into one field~\cite{Prokushkin:1998vn}:
\[
    C(y, K, \psi_{1,2}|x) = C^{\text{dyn}}(y, K, \psi_1|x)\psi_2 + C^{\text{aux}}(y, K, \psi_1|x),
\]
that transforms as the adjoint representation under gauge Lie algebra:
\begin{equation}
    \begin{aligned}
        D C(y, K, \psi_{1,2}|x) =& \; d C(y, K, \psi_{1,2}|x)\\
        &\; + W(y, K, \psi_1)\wedge C(y, K, \psi_{1,2}) - C(y, K, \psi_{1,2})\wedge W(y, K, \psi_1).
    \end{aligned}
\end{equation}
Here $C^{\text{dyn}}$ contains the first two cases of matter fields and $C^{\text{aux}}$ contains the last two.
While the former $C^{\text{dyn}}$, as matter fields, is well understood in many examples of 3d AdS/CFT duality,
the latter $C^{\text{aux}}$ behaves more exotic in aspects of representation theory, and have got no clear understanding in AdS/CFT correspondence so far.
In the other side, theories with only $C^{\text{dyn}}$ suffer from quantization problems beyond linear order.
Luckily, up tp linearized level every field is free and only interact with background field. Therefore, one can set $C^{\text{aux}} = 0$ consistently in linearized analysis around vacuum, 
which is enough for our purpose in this paper.

Now consider E.o.Ms of $C^{\text{dyn}}$ in pure AdS backgrounds that without any higher-spin background fields.
Here
\begin{equation}
    W^{\text{gr}}_\mu(y, K, \psi|x) = \frac{1}{8\mathrm{i}}(\omega_\mu^{\alpha\beta} + \lambda e^{\alpha\beta}\psi )y_{\alpha}y_{\beta},
\end{equation}where $\lambda$ is the cosmological constant.
With this connection, the covariant derivative $D$ of matter field $C^{\text{dyn}}(y, K, \psi|x)$ can be expressed as~\cite{Prokushkin:1998bq}:
\begin{equation}\label{eq:unfolded eom}
    D_\mathrm{L} C^{\text{dyn}}(y, K, \psi|x) = \lambda \{e^{\alpha\beta}y_{\alpha}y_{\beta}, C^{\text{dyn}}(y, K, \psi|x)\}
\end{equation}with $D_{\mathrm{L}}$ denotes the covariant derivative of local Lorentz connection $\omega_\mu^{\alpha\beta}$.

Expand $C(y, K, \psi|x)$ as 
\begin{equation}
    C(y, K, \psi|x) = \sum_{n=0}^\infty\sum_{I,J=0,1}\frac{1}{n!}C_{IJ}^{\alpha_1\dots\alpha_n}(x){(K)}^I y_{(\alpha_1}\dots y_{\alpha_n)}{(\psi)}^J,
\end{equation}identity Eq.~\eqref{eq:unfolded eom} is independent to $\psi$, allowing one to ignore the $\psi$-dependence. 
With the analysis of integrability condition for unfolded E.o.M, lowest components of $C^{\text{dyn}}(x)$ satisfy that~\cite{Prokushkin:1998bq, Barabanshchikov:1996mc}:
\begin{equation}
    D^\mu D_\mu C_I(x) {(K)}^I = \left(\frac{3}{2}\lambda^2 - \tilde{M}^2\right) C_I(x) {(K)}^I,
\end{equation}
\begin{equation}
    {e^\mu}_{\beta}^{\;\;\alpha} D_\mu C^\beta_I(x) {(K)}^I \equiv {\slashed D}_{\beta}^{\;\;\alpha} C^\beta_I(x) {(K)}^I = \frac{\tilde{M}}{\sqrt{2}} C^\alpha_I(x) {(K)}^I.
\end{equation}
where parameter $\tilde{M}^2$ relates to the element $K$. By the projection $P^K_{\pm}$ to subspace $C^{\pm} = P^K_{\pm}C $, one may replace $K$ by $\pm_K$,
and the mass parameter is~\cite{Prokushkin:1998bq}
\begin{equation}
    \tilde{M}^2_{\pm_K} = \lambda^2 \frac{\nu(\nu\mp_K 2)}{2} \Longrightarrow M^2 = \lambda^2\frac{(\nu\pm 1)(\nu\mp_K 3)}{2}, n\text{  even},
\end{equation}
\begin{equation}
    \tilde{M}_{\pm_K} = \mp_K\frac{\lambda \nu}{\sqrt{2}} \Longrightarrow M = \mp_K\lambda\nu/2, n\text{  odd},
\end{equation}correspond to bosons and fermions respectively. Here we redefine the mass parameter in original papers into a more straight one, by replacement
\begin{equation}
        M^2 =\tilde{M^2} - \frac{3}{2}\lambda^2, \quad \text{bosonic case, }n\text{  even},
\end{equation}
\begin{equation}
    M = \frac{\tilde{M}}{\sqrt{2}}, \quad \text{fermionic case, }n\text{  odd}.
\end{equation}

In each subspace under projection $P^K$ and $P^\psi$, lowest-spin coefficients $C(x)$ and $C_{\alpha}(x)$ giving the boson and fermion freedom while coefficients with higher spin are decided by constraints from self-consistency.

In conclusion, matter field components of $\cn = 2$ Vasiliev higher-spin theory can be listed as Table.\ref{tab:matter_field}
\begin{table}[H]
    \centering
    \setlength{\tabcolsep}{16pt}
    \renewcommand{\arraystretch}{2}
    \begin{tabular}{c|c|c}
        mass $M$ & $K = +$ & $K = -$ \\
        \hline
        $\pi_f=1$   & $M^2_+ = \lambda^2 \frac{(\nu + 1)(\nu - 3)}{2}$  & $M^2_- = \lambda^2 \frac{(\nu + 3)(\nu - 1)}{2}$ \\
        \hline
        $\pi_f=-1$  & $M_+ = -\lambda \nu$  & $M_- = \lambda \nu$ \\
    \end{tabular}
    \caption{Matter field components of both $\psi = \pm$ subspace and their mass, distinguished by eigenvalue of $\pi_f$ and $P^K_\pm$.}
    \label{tab:matter_field}
\end{table}
Where there are two bosons and two fermions in both $\psi = \pm$-subspace, each forming one $N = 2$ supermultiplet with respect to diagonal $\cn = 2$ superalgebra generated by $Q^1_\alpha \propto y_{\alpha}, Q^2_\alpha \propto y_\alpha K$.
By simply regarding all coefficient function $C_{IJ}^{\alpha_1\dots\alpha_n}$ as complex functions, these supermultiplet becomes complex.
For every $\cn = 2$ supermultiplet one have two choise of boundary conditions, giving different 1-loop partition function in bulk.
In the higher-spin gravity theory that are argued to be dual with $\cn = (2, 2)$ minimal models, the matter field contents are two such complex $\cn = 2$ supermultiplets, with both the boundary conditions put on each of them separately.
\subsubsection{Bosonic projection}
By projecting $\mathrm{Aq}(2, \nu)$ into its $\pi_f = 1$ subspace one get an bosonic higher-spin algebra and the corresponding trace. 
Traceless part of this associative algebra also form a Lie algebra, isomorphic to the $\mathrm{hs}[(1-\nu)/2]\oplus \mathrm{hs}[(1+\nu)/2]\oplus \mathrm{u}(1)$ Lie algebra.
Which is just the bosonic part of $\mathrm{shs}[(1-\nu)/2]$ Lie superalgebra. Each $\mathrm{hs}[\mu]$ factor giving a tower of higher-spin fields with spin $s=2,3,4\dots$, while $\mathrm{u}(1)$ becomes a spin-1 gauge field.
Matter fields values in the associative algebra above and contains only one boson fields in total, with mass parameter $M^2 = \lambda^2\frac{(\nu\pm 1)(\nu\mp 3)}{2}$.

\section{An \texorpdfstring{$\cn = (0, 2)$}{N = (0, 2)} generalization of linearized Vasiliev higher-spin supergravity}\label{sec:construction}
Now we discuss the generalization of constructions above into the $\cn = (0, 2)$ version with projected supersymmetry.

The key point of construction is to notice the holomorphic decomposition for both the structure of gauge algebra and the representation theory for it on matter fields, 
with which we can modify data of fields in one $\psi = \pm$-subspace while keep the symmetry structure of the other one.
\subsection{Gauge fields and asymptotic symmetry}
\subsubsection*{Gauge fields content}
Recall that in the construction shown in the previous section, 
gauge fields divides naturally into two parts by the direct plus structure of the gauge algebra $\mathrm{shs}[(1-\nu)/2] \oplus \mathrm{shs}[(1-\nu)/2]$, represented by $\psi = \pm$, 
while for each $\psi = \pm$ part of the matter field $C^{\text{dyn}}$ selected, 
$\psi = \pm$ part of gauge (higher-spin) fields act on it from separate sides. 
Therefore, we can separately modify data of fields in each single $\psi = \pm$ part without influence to the other part.

Now for gauge fields, we replace the right-moving part, in other word the $\psi = +$ part of the gauge Lie algebra ${\mathrm{shs}[(1-\nu)/2]}_{\text{R}}$ of $N = 2$ Vasiliev higher-spin theory into its bosonic subalgebra, which is the $\pi_f = 1$ subspace $\mathrm{hs}[(1-\nu)/2]\oplus \mathrm{hs}[(1+\nu)/2]\oplus \mathrm{u}(1)$.
Since the latter Lie algebra is a subalgebra of the former, any representation of the former $\mathrm{shs}[(1-\nu)/2]$ is automatically a (reducible) representation of ~\cite{Creutzig:2011fe} the latter $\mathrm{hs}[(1-\nu)/2]\oplus \mathrm{hs}[(1+\nu)/2]\oplus \mathrm{u}(1)$. 

Since the injection of Poincare subalgebra $\mathrm{so}(2, 2)$ into the left and right factor of gauge Lie algebra have not been changed by this bosonic projection, this new Lie algebra is still available to be the gauge algebra of the higher-spin and gravity part of our higher-spin gravity theory. Now
our theory contains one supersymmetric higher-spin tower and one bosonic higher-spin tower, with gauge fields of spin $(s, s+1/2,s+1/2, s+1) \oplus (s, s+1)$, $s = 1, 2, 3, \dots$, which is listed in Table.\ref{tab:field_components}.
\subsubsection*{Asymptotic symmetry}
One way to demonstrate that such a theory may correspond to $\cn = (0, 2)$ CFT in $2\mathrm{d}$ is to write down its asymptotic symmetry, on the background of AdS vacuum. Asymptotic symmetries of bulk theory is believed to correspond to global symmetries of the dual boundary theory.
If structure of $\cn = (0, 2)$ supersymmetry is found in the algebra of asymptotic symmetries of our theory, its dual theory, if exists, should have the same $\cn = (0, 2)$ supersymmetry, as is desired.

Concretely, gauge symmetries in the bulk and asymptotic symmetries is linked by the so-called ``Drinfeld-Sokolov reduction'', a hamiltonian reduction in highest-weight gauge~\cite{Coussaert:1995zp,Henneaux:1999ib} for not only gravity but also higher-spin case~\cite{Henneaux:2010xg,Campoleoni:2010zq,Campoleoni:2011hg}. It is important that the reduction is done separately for each $\psi = \pm $ part of the gauge symmetry. 
To clarify that, notice that $\psi = \pm$ part of gauge fields are respectively represented by $A$ and $\bar{A}$ in Chern-Simons formalism. Because of the direct plus structure of gauge symmetry algebra, both $A$ and $\bar{A}$ satisfy gauge invariance separately. Therefore, the calculations of asymptotic symmetries can be done separately for them, as is the case in ~\cite{Campoleoni:2011hg, Creutzig:2011fe}.

Therefore, with the conclusions that have already been known for long~\cite{Henneaux:2010xg, Campoleoni:2011hg, Creutzig:2011fe}, the ${\mathrm{shs}[(1-\nu)/2]}_{\text{L}}$ and ${\mathrm{hs}[(1-\nu)/2]\oplus \mathrm{hs}[(1+\nu)/2]\oplus \mathrm{u}(1)}_{\text{R}}$ gauge symmetry corresponds to ${\mathrm{sW}_{\infty}[(1-\nu)/2]}_{\text{L}}$ and ${\left(\mathrm{W}_{\infty}[(1-\nu)/2]\oplus \mathrm{W}_{\infty}[(1+\nu)/2]\oplus \mathrm{u}(1)\right)}_{\text{R}}$ 
respectively\footnote{The Drinfeld-Sokolov reduction mentioned here is the ``classical'' one, meaning that it is acted on some classical Poisson algebra.
Therefore the resulted W-algebra do not contains quantum correction terms, but only some semi-classical ($c \rightarrow \infty $) limit of those quantum W-algebra that of finite $c$.}
, in where there contains the subalgebra of $2\mathrm{d}\; \cn = (0, 2)$ supersymmetry, making it reasonable to call such a theory that to be constructed as an $\cn = (0, 2)$ higher-spin gravity theory.

\subsection{Matter fields}
Matter fields decomposes into different parts accordingly,
where the original irreducible $N = 2$ supermultiplet branches into two supermultiplets of the $\cn = 2$ superalgebra in $\psi = \pm$-subspace left, as is shown in Table.\ref{tab:q-action}.

To deduce the result, recall that in each $\psi = \pm$ subspace of matter field, both the two $\psi = \pm$ subspace of gauge Lie algebra act only from left or right on matter field, while algebra within each $\psi = \pm$ subspace choose the opposite sides.
That is to say, when considering the $\psi = +$ subspace of matter fields, action of $\psi = +$ subalgebra is from the left side, while $\psi = -$ subalgebra from the right side. The opposite happens for the $\psi = -$ subspace.
Therefore, when supersymmetries of one of $\psi = \pm$ subspaces is removed, both $\psi = \pm$ subspace are acted by supercharges from only one side, therefore branches into two smaller supermultiplets.
In each $\psi = \pm$ part, these two supermultiplets are:
\begin{itemize}
    \item one real boson with $M^2 = \lambda(\nu+1)(\nu-3)/2$ and one Majorana fermion with $M = \pm\lambda\nu/2$. We call it $[\psi=\pm, K=+]$ supermultiplet, since the bosonic part lies in $K=+$ subspace.
    \item one real boson with $M^2 = \lambda(\nu-1)(\nu+3)/2$ and one Majorana fermion with $M = \mp\lambda\nu/2$. we call it $[\psi=\pm, K=- ]$ supermultiplet, since the bosonic part lies in $K=-$ subspace.
\end{itemize}It seems exotic that the majorana fermions have both signature of mass parameter. In fact, as is discussed in section 4.1 of ~\cite{Creutzig:2011fe}, when one put a Dirac fermion $\psi$ into Euclidean space-time, $\psi$ and $\bar{\psi}$ can be treat as two independent Majorana fermion
with opposite signature of mass parameter.
Therefore, the subtlety of signature require us not to treat these fermions as Dirac fermions. 
From another perspective, in the ``unfolded formalism'' each supermultiplets contain only one of the $K=+$ or $K=-$ term of $C^{\text{dyn}}(y,K,\psi|x)$, 
while complex conjugation brings additional $(-1)$ factor
\[
y_{\alpha_1}\dots y_{\alpha_n}K \rightarrow {(-1)}^n y_{\alpha_1}\dots y_{\alpha_n}K
\]
in order to be compatible with the relation $\{K, y_\alpha\} = 0$,
transforming fermions in $K = +$ subspace into $K = -$ subspace.
One may analogize this signature to the chirality of fermions in even dimensions,
by which fermions in these supermultiplets is analogous to ``Weyl-Majorana'' fermions in $2\mathrm{d}$.

\begin{table}[h]
\centering
\setlength{\tabcolsep}{16pt}
\renewcommand{\arraystretch}{5}
\begin{tabular}{c|c c}
    Matter fields & $K = +$ & $K = -$  \\
    \hline
    $\pi_f = 1$ & $\tikzmarknode{cell11}C^+$\!$\tikzmarknode{cell11'}(x)$\! $\tikzmarknode{cell11''}P^K_+$ & $\tikzmarknode{cell12}C^- (x)$\! $\tikzmarknode{cell12'}P^K_-$  \\
    $\pi_f = -1$ & $C^{+,\alpha}$\! $\tikzmarknode{cell21}(x)$\! $\tikzmarknode{cell21'}y_\alpha$\! $\tikzmarknode{cell21''}P^K_+$ & $\tikzmarknode{cell22}C^{-,\alpha} (x)$ \!$\tikzmarknode{cell22'}y_\alpha$ \!$\tikzmarknode{cell22''}P^K_-$ \\
\end{tabular}
\caption{The $Q$-action on matter fields of Vasiliev higher-spin gravity. The blue arrows refer to left actions while green arrows refer to right ones. 
It's easy to see that for adjoint representation, as both side of action is considered, these four fields constructs one $\cn = 2$ supermultiplet, 
while when considering pure left or right action, as is the case of considering action of single $\psi = \pm$-subspace of the gauge algebra, 
this supermultiplet splits into two supermultiplets, each containing one boson and one fermion field, as is easy to read from the table.}
\label{tab:q-action}
\end{table}
\begin{tikzpicture}[remember picture, overlay]

\draw[blue, -{Stealth[length=3mm]}, thick] 
    (cell11) --
    (cell21)
    node[midway, left] {$Q^+$};

\draw[blue, -{Stealth[length=3mm]}, thick] 
    (cell21') --
    (cell11')
    node[midway, right] {$Q^-$};

\draw[blue, -{Stealth[length=3mm]}, thick] 
    (cell12') --
    (cell22'') 
    node[midway, right] {$Q^-$};

\draw[blue, -{Stealth[length=3mm]}, thick] 
    (cell22') --
    (cell12)
    node[midway, left] {$Q^+$};

\draw[green!60!black, -{Stealth[length=3mm]}, thick] 
    (cell11'') to[bend left=15]
    node[pos=0.3, above] {$Q^-$}
    (cell22);

\draw[green!60!black, -{Stealth[length=3mm]}, thick] 
    (cell22) to[bend left=15]
    node[pos=0.7, below] {$Q^+$}
    (cell11'');

\draw[green!60!black, -{Stealth[length=3mm]}, thick] 
    (cell21'') to[bend left=15]
    node[pos=0.7, above] {$Q^-$}
    (cell12);

\draw[green!60!black, -{Stealth[length=3mm]}, thick] 
    (cell12) to[bend left=15]
    node[pos=0.3, below] {$Q^+$}
    (cell21'');

\end{tikzpicture}

Now, matter fields with higher-spin gauge fields taking value in 
\[
{\mathrm{shs}[1-\nu]}_{\text{L}} \oplus {\left(\mathrm{hs}[1-\nu]\oplus \mathrm{hs}[1+\nu]\oplus \mathrm{u}(1)\right)}_{\text{R}}
\] should be built by these four supermultiplet.
All of them have the structure of $\cn = (0, 2)$ short supermultiplet in $2\mathrm{d}$ CFTs.

Field components of our $\cn = (0, 2)$ higher-spin gravity theory is listed in Table.\ref{tab:field_components}.
\begin{table}[H]
    \centering
    \renewcommand{\arraystretch}{1.5}
    \begin{tabular}{c|c|c}
             & $\psi=+$ & $\psi=-$ \\
        \hline
        $\cdots$ & $\cdots$ & $\cdots$ \\
        high $s $ & spin=$(s, s+1)$ & spin=$(s, s+1/2, s+1/2, s+1)$ \\
        $\cdots$ & $\cdots$ & $\cdots$ \\
        high $s = 2$ & spin=$(2, 3)$ & spin=$(2, 5/2, 5/2, 3)$ \\
        high $s = 1$ & spin=$(1, 2)$ & spin=$(1, 3/2, 3/2, 2)$ \\
        \hline
        low $s < 1$ & $[\psi=+,K=\pm]$ & $[\psi=-,K=\pm]$\\
    \end{tabular}
    \caption{Field components of high and low spin $s$. $(s_1, s_2, s_3, \dots)$ means there exists higher-spin massless fields with each spin $s_i$, while $[\psi=\pm,K=\pm]$ represents four possible kinds of matter supermultiplets.}
    \label{tab:field_components}
\end{table}
\section{One-loop partition function}\label{sec:one_loop_partfunc}
One-loop partition functions of such a $\cn = (0, 2)$ higher-spin gravity theory with different, around the thermal AdS vacuum, can be calculated by the heat-kernel method~\cite{Giombi:2008vd, David:2009xg,Creutzig:2011fe}.

Since that in the thermal AdS vacuum only background field of gravity is non-zero, and that in linear theory and up to 1-loop order, gauge fields and matter fields couples to only these background fields, 
contributions of every fields to the 1-loop partition functions can be dealt with separately, as contributions of free fields propagating in the AdS background. 
Among these fields, 1-loop partition functions of matter fields and those gauge fields that be able to transform into Fronsdal (metric-like) form, which is equavalent to say it appears in both $\psi = \pm$ part, is already known~\cite{Giombi:2008vd,Creutzig:2011fe}.

For matter fields,
\begin{equation}\label{eq:boson_partfunc}
    Z_{\text{boson}}^h = \prod_{i,j=0}^{\infty}\frac{1}{1-q^{i+h}{\bar{q}}^{j+h}}, h = \frac{1}{2}(1+\sqrt{1+M^2});
\end{equation}
\begin{equation}\label{eq:dirac_partfunc}
    Z_{\text{Dirac}}^h = \prod_{i,j=0}^{\infty}\big(1+q^{i+h}{\bar{q}}^{j+h+1/2}\big)\big(1+q^{i+h+1/2}{\bar{q}}^{j+h}\big), h = \frac{1+M}{2};
\end{equation}
For gravity and higher-spin fields~\cite{Gaberdiel:2010ar, Creutzig:2011fe},
\begin{equation}\label{eq:boson_hs_partfunc}
    Z_{\text{HS}}^s = \prod_{n = 0}^\infty \frac{1}{(1-q^{n+s})}\frac{1}{(1-{\bar{q}}^{n+s})}, s\in \mathbb{N} \text{ and } s\geq 2;
\end{equation}
\begin{equation}\label{eq:fermion_hs_partfunc}
    Z_{\text{HS}}^s = \prod_{n = 0}^\infty (1+q^{n+s})(1+{\bar{q}}^{n+s}), s\in \mathbb{N} + \frac{1}{2} \text{ and } s\geq 3/2.
\end{equation}
while partition function of those gauge fields that appearing in only one $\psi = -$ part cannot be transformed into metric-like form, therefore still remains to be calculated.
Besides, some subtlety about the signature of mass parameter appear in fermionic matters, exceeding the original calculation for Dirac fermions.

The detail of our calculation is shown in Sec.\ref{sec:calculation}, with results in line with expectations:
\begin{equation}\label{eq:majorana_chiral_partfunc1}
    Z_{\text{Majorana}}^{h,\zeta=+} = \prod_{i,j=0}^\infty \big(1+q^{i+h+1/2}{\bar{q}}^{j+h}\big), h = \frac{1+ M}{2};
\end{equation}
\begin{equation}\label{eq:majorana_chiral_partfunc2}
    Z_{\text{Majorana}}^{h,\zeta=-} = \prod_{i,j=0}^\infty \big(1+q^{i+h}{\bar{q}}^{j+h+1/2}\big), h = \frac{1+ M}{2};
\end{equation}
\begin{equation}\label{eq:fermion_chiral_hs_partfunc}
    Z_{\text{HS}}^{s,\psi=-} = \prod_{n=0}^\infty (1+{\bar{q}}^{n+s}), s\in \mathbb{N} + \frac{1}{2} \text{ and } s\geq 3/2.
\end{equation}

\subsection{Calculation of 1-loop partition function}\label{sec:calculation}
The calculation of 1-loop partition function around thermal AdS$_3$ space-time with modular parameter $2\pi\tau = \theta + \mathrm{i}\beta$ is shown. 
Before any details we clarify several fundamental concepts and idea to be used later.

By thermal AdS$_3$ space-time, we means the quotient of global EAdS$_3$ space-time by isometry of group $\mathbb{Z}$. The global EAdS$_3$ space-time is discribed by coordinate $(t,r,\phi); t,r\in (-\infty, +\infty),\phi\in [0,2\pi)$, under which metric becomes:
\begin{equation}
    {\mathrm{d}s}^2/l^2_{\mathrm{AdS}} = (1+r^2){\mathrm{d}t}^2 + \frac{1}{1 + r^2}{\mathrm{d}r}^2 + r^2{\mathrm{d}\phi}^2
\end{equation} while it is quotient by the isometry $ t \rightarrow t + \beta; \phi \rightarrow \phi + \theta $.
It is easy to check that the boundary of such a space-time is just the $2\mathrm{d}$ torus with modular parameter $\tau$.

As is argued in the main body of this paper, with no other background fields opened, in the linearized action dynamical fields behave like free fields propagating on the AdS background and do not couple to each other, 
making their contribution to 1-loop partition function factorized:
\begin{equation}
    S_{\text{linearized}}[B_i, \Phi_j; \bar{\omega}, \bar{e}] = \sum_{\text{gauge fields}} S_{\text{linearized}}[B_i; \bar{\omega}, \bar{e}] + \sum_{\text{matter fields}} S_{\text{linearized}}[\Phi_j; \bar{\omega}, \bar{e}],
\end{equation}where $\bar{\omega}, \bar{e}$ represents the background local Lorentz connection and frame fields of AdS space-time; linearized action of gauge (higher-spin) fields will be determined by linearization of the Chern-Simons action 
that is used to describe the interaction between higher-spin fields:
\begin{equation}
    Z = \exp \left( {iI[{A^{(0)}}]} \right)
    \int {\left[ {D B} \right]\exp \left( {\frac{k}{{4\pi }}\int_M {{\epsilon ^{\mu \nu \rho }}B_{\mu, \alpha_1\cdots \alpha_{2s-1}} D_\nu {B_\rho}^{\alpha_1\cdots \alpha_{2s-1}}} } \right)},
\end{equation}where $A = B + A^{(0)}$ and $A^{(0)}$ is the background field and $D_\mu$ covariant derivative with background connections; linearized actions of matter fields, as whose E.o.M. is determined by the ``unfolded formalism'' and reduced to E.o.M. of free fields, is just actions of free bosons and fermions
with corresponding mass parameter that propagating on AdS background:
\begin{equation}
    Z_{\text{boson}} = 
    \int {\left[ {D \phi} \right]\exp \left( \int_M {\sqrt{-g}\phi (\nabla^2 - M^2) \phi} \right)},
\end{equation}

\begin{equation}
    Z_{\text{fermion}} = 
    \int {\left[ {D \psi} \right]\exp \left( \int_M {\sqrt{-g}\psi^\alpha {(\slashed\nabla - M)}_\alpha^{\;\,\beta} \psi_\beta} \right)},
\end{equation}

Thus, factors of 1-loop partition functions of listed in Table.\ref{tab:field_components} can be calculated separately. As is said in the main body, what remains to calculate is those half-integer spin higher-spin fields that appear in $\psi=-$ only, and fermions with certain choice of signature of mass parameter $M$.
For both cases we will deal with the path integral by Faddev-Popov method and deduce a functional determinant form of partition function, then calculate these determinants by heat-kernel method. That's to say, we are following the strategy of ~\cite{David:2009xg,Creutzig:2011fe}.

In the remain part, these contributions to the 1-loop partition function of every field are directly called as ``partition function'' for short.
\subsubsection{Functional determinant form of partition function}

\noindent\textbf{higher-spin fields}

Half-integer spin higher-spin fields in $\psi=-$ satisfy the linearized gauge invariance:
\begin{equation}
    \delta B_{\mu}^{\;\;\alpha_1\dots \alpha_{2s-1}} = D_\mu \epsilon^{\alpha_1\dots \alpha_{2s-1} } - (\psi) \frac{2s-1}{2}{{\bar{e}}_{\mu,\beta}}^{(\alpha_1|} \epsilon^{\beta |\alpha_2\dots \alpha_{2s-1})},
\end{equation}as $\psi=-$ here. Still, to make the calculation general we keep the signature $\psi$ all along the calculation below.
Free field action in AdS background is uniquely determined by invariance under this gauge transformation~\cite{Creutzig:2011fe} as the linearized Fang-Fronsdal action of fermionic higher-spin field~\cite{Fang:1978wz,Fronsdal:1978rb}. Following~\cite{Creutzig:2011fe}, path integral results in the functional determinant:
\begin{equation}
    Z_{\text{HS}}^{s,\psi} = \frac{{{{\det }^{1/2}}\left( {\slashed{\nabla } - \psi (s - 1)} \right)_{\left( s \right)}^{{\rm{TT}}}}}{{{{\det }^{1/2}}\left( {\slashed{\nabla } - \psi s} \right)_{\left( {s - 1} \right)}^{{\rm{TT}}}}},
\end{equation}where $\mathrm{TT}$ means that the operators acts on traceless transversal fields that corresponds to physical d.o.f. of massless fields.

\noindent\textbf{Majorana fermion}

The path integral for Majorana fermions with certain mass parameter can be directly integrated out as:
\begin{equation}
    Z_{\text{Majorana}}^{h,\zeta} = {{\det }^{1/2}}\left( {\slashed{\nabla } - \zeta M} \right)
\end{equation}
The main problem now is to calculate these determinants of slashed derivative $\slashed{\nabla} + \cdots$.
\subsubsection{Heat-kernel method}
Functional determinants can be calculated by a relative heat-kernel that collecting all information of eigenstates of the operator inside the determinant.

\noindent\textbf{Heat-kernel}

Consider a normal derivative operator $D$ whose eigenvalue $\lambda$ all have positive real part $\mathrm{Re}\lambda > 0$ acting on some function space, we can define its heat-kernel
\begin{equation}
    {K_{D}}(x,a,y,b;t) = \sum_\lambda \phi_\lambda(x,a) \phi^*_{\lambda}(y,b)e^{-\lambda t},
\end{equation}with $\phi_\lambda(x,a)$ the corresponding eigenstates and $a, b$ inner d.o.f. It is the Green function of heat equation
\[
    {\left(\frac{\partial}{\partial t}+D\right)} f(t;x,a) = \phi^B(x,a).
\]

As the heat-kernel collects all information of eigenstates, the functional determinant of $D$ can be represented by the trace of heat-kernel $K_D(t) = \int\mathrm{d}x^n\sum_{a} {K_{D}}(x,a,x,a;t)$ as~\cite{David:2009xg}
\begin{equation}
    \ln \det D =\sum_{\lambda} \ln\lambda = \int_{0}^{\infty}\frac{\mathrm{d}t}{t} \sum_\lambda e^{-\lambda t} = \int_{0}^{\infty}\frac{\mathrm{d}t}{t} K_{D}(t).
\end{equation}
In addition, since the thermal AdS space-time is a quotient, we can express its trace as~\cite{Gopakumar:2011qs,Creutzig:2011fe}:
\begin{equation}
    K_{D}(t) = \sum_{\gamma^m\in\mathbb{Z}} {(-)}^{\mathrm{F}}\int_{M/\mathbb{Z}} \mathrm{d}x \sum_\lambda \phi^A_\lambda(x)\phi^*_{\lambda A}(\gamma^m (x)) e^{-\lambda t}.
\end{equation}where $M$ is chosen to be the global Euclidean AdS space-time. $\mathrm{F}$ is the fermion parity and ${(-)}^{\mathrm{F}}$ represents the fermionic statistics, as is shown in~\cite{Creutzig:2011fe}.

Thus, the calculation of heat-kernel is transformed into an eigenvalue problem.
To solve the eigenvalue problem, a representation theory method~\cite{Camporesi:1995fb} is used.

\noindent\textbf{Eigenvalue problem with aspect to representation theory}

As a maximally symmetric space-time, the global Euclidean space-time can be expressed as a coset space~\cite{David:2009xg}:
\begin{equation}
    \mathrm{EAdS}_3 \cong \mathrm{SO}(3,1)/\mathrm{SO}(3)\cong \mathrm{Spin}(3,1)/\mathrm{Spin}(3),
\end{equation}where $\mathrm{Spin}(3)$ is just the universal cover of (Euclidean) Lorentz group, that all fields should be a representation of it.
An important property is satisfied by this coset space representation, that one can identify the orthogonal complementary space of corresponding Lie algebra $\mathrm{so}(3)$ in $\mathrm{so}(3,1)$ with respect to the Killing form, denoted as $P$ for their physical meaning as translation generators, that satisfy
\begin{equation}
    [P, P] \in \mathrm{so}(3),
\end{equation}
which by definition makes the space-time a Riemannian symmetric space, ensuring the compatibility between Riemannian structure and group action on this coset space, that $P$ can be identified as parallel transports along geodesics generated by corresponding tangent vectors and $\mathrm{sp}(3)$ regarded as local Lorentz transformations.
The detail can be found in section 5.3 of \cite{Camporesi:1995fb}.

Now consider a representation $R$ of $\mathrm{Spin}(3)$ and some field $\phi^A(x)$ that transforms as this representation.
$\phi^A(x)$ can be regarded as a section of some vector bundle, with each fiber a representation space $V^R$ and the construction group be the local $\mathrm{Spin}(3)$ transformation group.
In a more mathematical language, this local $\mathrm{Spin}(3)$ transformation is described by a principal $\mathrm{Spin}(3)$ bundle, which is just homeomorphic to the $\mathrm{Spin}(3,1)$ as a manifold, and our vector bundle is an adjoint bundle of this principal bundle.
The configuration space of $\phi^A(x)$ is certainly the section space of this adjoint bundle.
But since our space-time, the base manifold, is a coset space $\mathrm{Spin}(3,1)/\mathrm{Spin}(3)$ with $\mathrm{Spin}(3)$ in the denominator, 
The section space of adjoint bundle with representation $R$ of $\mathrm{Spin}(3)$ on this manifold actually form an induced representation $\mathrm{ind}_{\mathrm{Spin}(3)}^{\mathrm{Spin}(3,1)}(R)$
\footnote{More concretely, with $\psi^A$ covariant under $\mathrm{Spin}(3)$, any section of the adjoint bundle can be described as~\cite{David:2009xg}
\begin{equation}
    {\psi^i(g)} = \psi^i(g_0g_0^{-1}g) = {R(g^{-1} g_0)}^i_{\ j}\psi^i([g])
\end{equation}
where $[g]\in \mathrm{Spin}(3, 1)/\mathrm{Spin}(3)$ is a coset and is used to denote points of Euclidean AdS space-time; $g_0([g])$ serves as a selected section of the principle $\mathrm{Spin}(3)$ bundle, 
in the sense that $g_0 \in \mathrm{Spin}(3, 1)$ in each $[g]$ is regarded as a point of the principal bundle by homeomorphism between this principal $\mathrm{Spin}(3)$ bundle and  $\mathrm{Spin}(3, 1)$.
Now as a ($\mathrm{Spin}(3)$-equivariant) function on $\mathrm{Spin}(3, 1)$ manifold, ${\psi^i (g)}$ immediately forms a representation of $\mathrm{Spin}(3, 1)$.
}
which can be decomposed into different irrep.\ of the Poincare group $\mathrm{Spin}(3,1)$.

In fact, the decomposition of induced representation is given by the Frobenius reciprocity theorem, which says that:
\begin{equation}\label{Frobenius_reciprocity_theorem}
    \mathrm{mult}_G[\mathrm{ind}_H^G(R), S] = \mathrm{mult}_H[S, R],
\end{equation}
where $R$ is an $H$-irrep.\ while $R$ is a $G$-irrep. $\mathrm{mult}_G[R_1, R_2]$ represents the multiplicity of irrep.\ $R_2$ appearing in the decomposition of $R_1$ as $G$ representations.

What counts is that, each $H$-irrep.\ $S$ in the decomposition of $\mathrm{ind}_H^G(R)$ is an eigenspace of Laplacian, with eigenvalues related to the Casimir operator of $G$ and $H$ group~\cite{Camporesi:1995fb} as:
\begin{equation}
    -E = C_2^G(R) - C_2^H(S),
\end{equation}simplifying the eigenvalue problem into a counting problem of representations. 
Eigenvalue problems of Dirac operators, as ${\slashed\nabla}^2 = \Delta + \text{scalars}$, should also simplify greatly therefore.

\subsubsection{Calculate heat-kernel of Dirac operators}
In this subsection, following \cite{David:2009xg, Gopakumar:2011qs}, we will transform between these three manifold many a time: sphere $S^3 \cong \mathrm{Spin}(4)/\mathrm{Spin}(3)$, 
Euclidean AdS space-time $\mathrm{EAdS}_3 \cong \mathrm{Spin}(3, 1)/\mathrm{Spin}(3)$ and (Minkowskian) AdS space-time $\mathrm{AdS}_3 \cong \mathrm{Spin}(2, 2)/\mathrm{Spin}(2, 1)$, 
all tof them are coset spaces and can be wick rotated to each other. The ``wick rotation'' between $S^3$ and $\mathrm{EAdS}_3$ is done by~\cite{David:2009xg}
coordinate transformation
\[
    r \rightarrow \sinh\rho, t\rightarrow t, \phi\rightarrow\phi
\]and then continuation
$\mathrm{i}t\rightarrow\eta, -\mathrm{i}\rho\rightarrow\psi$ with $ {\mathrm{d}s}_{\mathrm{EAdS}}^2\rightarrow - {\mathrm{d}s}_{\mathrm{S^3}}^2 $.
Notice that since the signature of ${\mathrm{d}s}^2$ is changed, so does all eigenvalues of Laplacian during this continuation.

On $\mathrm{S}^3$ manifold, each symmetric traceless transverse (STT for short) field with spin-$s$ forms an induced representation of $\mathrm{Spin}(4)$
that decomposes into $\mathrm{Spin}(4)\cong \mathrm{Spin}(2)\times \mathrm{Spin}(2)$ irrep.\ as
\begin{equation}
    \lambda_+ = (\frac{n}{2} + s, \frac{n}{2}); \lambda_- = (\frac{n}{2}, \frac{n}{2}+s).
\end{equation} with eigenvalues of Laplacian take
\begin{equation}
    -E_n^{(s)} = 2(C_2(\frac{n}{2}+s)+C_2(\frac{n}{2})) - C_2(s) = (s+n+2)(s+n) - s,
\end{equation} by analytic continuation, for EAdS, we have $\mathrm{Spin}(3, 1)\cong \mathrm{SL}(2, \mathbb{C})\cong \mathrm{SL}(2)\times \overline{\mathrm{SL}(2)} $ irrep.\ :
\begin{equation}\label{rep_decompos}
    \lambda_+ = (\frac{s - 1+ \mathrm{i}\lambda}{2}, \frac{-s - 1+ \mathrm{i}\lambda}{2}), \lambda_- = (\frac{-s - 1+ \mathrm{i}\lambda}{2}, \frac{s - 1+ \mathrm{i}\lambda}{2}), \lambda\in \mathbb{R}_+
\end{equation}where $n$ and $\lambda$ are related as $n\leftrightarrow -s-1+\mathrm{i}\lambda $ while continuation, with eigenvalue
\begin{equation}
    -E_\lambda^{(s)} = (\lambda^2 + s + 1).
\end{equation}
In fact, the heat-kernel of Laplacian can be directly calculated by data above, as is done in~\cite{David:2009xg} and some other works. However, as what is concerned here is the Dirac operator, further analysis still remains to be done.

Consider the Dirac operator ${\slashed\nabla}_{(s)} + M$ acting on a STT field with spin $s$, whose representation decomposition has been gotten as Eq.~\eqref{rep_decompos}.
Since ${\slashed{\nabla}_{(s)}}^2 = \Delta_{(s)} + (s-1)$, square of ${\slashed\nabla}_{(s)}$ is diagonalized by acting on irreps.
In addition, it takes the same eigenvalues on representation space labelled by $\lambda_\pm$ for each given $\lambda$. 
Therefore, the action of ${\slashed\nabla}_{(s)}$, and therefore ${\slashed\nabla}_{(s)} + M$, is closed in subspace spanned by irrep.\ $\lambda_+$ and $\lambda_-$ for each $\lambda$.

Now discuss the behavior of ${\slashed\nabla}_{(s)} + M$ in each of such subspaces. Though Euclidean is fine, we wick rotate the space-time into (Minkowskian) AdS space-time, where the two factor of Poincare group
$\mathrm{Spin}(2, 2)\cong \mathrm{SL}(2, \mathbb{R})\times \mathrm{SL}(2, \mathbb{R})$ are independent. Generators in these two $\mathrm{SL}(2, \mathrm{R})$s are denoted as $L_a$ and $R_a$ respectively, 
with Poincare generators represented as 
\[
M_a = L_a + R_a, P_a = L_a - R_a.
\]
When acting on STT fields with spin $s$, Dirac operators can be written as
\begin{equation}
    \begin{aligned}
        &{\left(\slashed P \right)}_{(\alpha_1|}^{\;\;\;\beta} {\phi}_{\beta|\alpha_2 \dots \alpha_{2s}) } \\
         =&\; {(\gamma^a)}_{({a_1}|}^{\;\;\;\;\beta}{\nabla_{a}}{\phi}_{\beta|\alpha_2 \dots \alpha_{2s}) }\\
         =&\; \frac{1}{2s}\sum_m {{(\gamma^a)}_{a_m}^{\;\;\;\beta}{\nabla_a}{\phi_{\beta{\alpha_1}\cdots\widehat {{\alpha_m}}\cdots\alpha_{2s}}}} \\
         =&\; \frac{1}{2s}{M^a}^{\left( {{\mathrm{spin - s}}} \right)}{P_a}^{\left( {{\mathrm{spin - s}}} \right)}{\left(\phi_{\alpha_1\alpha_2\cdots\alpha_{2s}}\right)} \\
         =&\; \frac{1}{2s}{\left( {L + R} \right)^a}^{\left( {{\mathrm{spin - s}}} \right)}{\left( {L - R} \right)_a}^{\left( {{\mathrm{spin - s}}} \right)}{\left(\phi_{\alpha_1\alpha_2\cdots\alpha_{2s}}\right)},
    \end{aligned}
\end{equation}
where $L^{\text{(spin-s)}}, R^{\text{(spin-s)}}$ means that they act on the spin $s$ STT field space. 
Now Dirac operators equals to $\frac{1}{2s}\left( {{L^2} - {R^2}} \right)$, acting diagonally on each irrep.\ of $\mathrm{Spin}(2, 2)$.
Denote the eigenvalue of Dirac operator $\slashed \nabla$ on representation $\lambda_\pm$ as $A_{\lambda_\pm}$.
Apply this relation to STT fields on $\mathrm{S}^3$, we get that:
\begin{equation}
    A_{\lambda_\pm} = \frac{{ \pm \sqrt{-1}}}{s}\left( {C_2\left( {\frac{n}{2} + s} \right) - C_2\left( {\frac{n}{2}} \right)} \right) =  \pm \sqrt{-1}\frac{{\left( {n + s + 1} \right)s}}{s} =  \pm \sqrt{-1}\left( {s + n + 1} \right),
\end{equation}where the additional $\sqrt{-1}$ come from the additional signature ${\mathrm{d}s}_{\mathrm{EAdS}}^2\rightarrow - {\mathrm{d}s}_{\mathrm{S^3}}^2 $ during analytic continuation.
Analytic continue this back to EAdS, one get eigenvalue data of STT spin $s$ fields:
\begin{equation}
        A_{\lambda_\pm}:
        \left( {s + n + 1} \right){\rm{i}}\left( {\begin{array}{*{20}{c}}
        1&0\\
        0&{ - 1}
        \end{array}} \right)\begin{array}{*{20}{c}}
        {{\lambda _ + }}\\
        {{\lambda _ - }}
        \end{array} 
        \longrightarrow
         \left( {\begin{array}{*{20}{c}}
        {{\rm{i}}\lambda }&0\\
        0&{ - {\rm{i}}\lambda }
        \end{array}} \right)\begin{array}{*{20}{c}}
        {\lambda_+}\\
        {\lambda_-}
        \end{array}
\end{equation}
which is compatible with eigenvalue data of Laplacian with 
\begin{equation}
    A^2_{\lambda_\pm} = E_\lambda^{(s)} + (s+1).
\end{equation}
Heat-kernel of Dirac operator itself can be calculated by these data of eigenvalues, 
with the Harish-Chandra character of $\mathrm{SL}(2, \mathbb{C})$~\cite{David:2009xg}:
\begin{equation}
    \chi_{(j_1, j_2)}(\alpha):= \chi_{(j_1, j_2)}(\mathrm{diag}(\alpha,\alpha^{-1})) = \frac{\alpha^{2j_1 + 1} {\bar{\alpha}}^{2j_2 + 1} + \alpha^{- 2j_1 - 1} {\bar{\alpha}}^{- 2j_2 - 1}}{{|\alpha - \alpha^{-1}|}^2},
\end{equation}
with $\mathrm{diag}(\alpha,\alpha^{-1})$ diagonal elements of $\mathrm{SL}(2, \mathbb{C})$, then
\begin{equation}
    \chi_{\lambda_\pm, s}(e^{\mathrm{i}m\tau/2}) := \frac{1}{2} \chi_{(\pm s-1+\mathrm{i}\lambda, \mp s-1+\mathrm{i}\lambda)}(\gamma^m) = \frac{\cos(\pm sm\tau_1 - \lambda m\tau_2)}{{|\sin m\tau/2|}^2}.
\end{equation}
giving the heat-kernel
\[
    \begin{aligned}
        & K_{\slashed{\nabla} + M}^{\left( s \right)}\left( {\tau ;t} \right)\\
        &= \frac{{{\tau _2}}}{{2\pi }}\sum\limits_{m \in \mathbb{Z} } {(-)}^{m}{\int_0^\infty  {{\rm{d}}\lambda } } 
        \left[ {{\chi _{{\lambda _ + }}}\left( e^{{{\mathrm{i}m\tau }}/{2}} \right) e^{ - \left( {{\rm{i}}\lambda  + M} \right)t} + {\chi_{{\lambda_- }}}\left( e^{{{\mathrm{i}m\tau }}/{2}} \right) e^{ - \left( { - {\rm{i}}\lambda  + M} \right)t}} \right]\\
        &= \sum\limits_{m \in \mathbb{Z}} \frac{{{(-)}^{m}{\tau _2}}}{{4\pi {{\left| {\sin \left( {m\tau /2} \right)} \right|}^2}}}
        {\int_{ - \infty }^\infty  {{\mathrm{d}}\lambda } } \cos \left( {ms{\tau_1} - m\lambda {\tau_2}} \right)  e^{ - \left( {{\rm{i}}\lambda  + M} \right)t}\\
        &= \sum\limits_{m \in \mathbb{Z}} \frac{{{{(-)}^{m}\tau _2}}}{{4\pi {{\left| {\sin \left( {m\tau /2} \right)} \right|}^2}}}
        {\int_{ - \infty }^\infty  {{\mathrm{d}}\lambda } } 
        \left( { e^{{{\rm{i}}ms{\tau_1} + {\rm{i}}\lambda ( - m{\tau _2} - t)}} + e^{ - {\rm{i}}ms{\tau _1} + {\rm{i}}\lambda (m{\tau _2} - t)}} \right)\frac{{e^{ - Mt}}}{2}\\
        &= \sum\limits_{m \in \mathbb{Z}} \frac{{{{(-)}^{m}\tau _2}}}{{2{{\left| {\sin \left( {m\tau /2} \right)} \right|}^2}}} 
        {\frac{1}{2}} e^{ - Mt}\left[ {{e^{{\rm{i}}ms{\tau _1}}}\delta \left( {-m{\tau_2} - t} \right) + {e^{ - {\rm{i}}ms{\tau _1}}}\delta \left( {m{\tau _2} - t} \right)} \right]\\
        &= \sum\limits_{m \geq 1} {\frac{{{(-)}^{m}{\tau _2}}}{{2{{\left| {\sin \left( {m\tau /2} \right)} \right|}^2}}}} 
        {e^{ - {\rm{i}}ms{\tau _1} - m M{\tau _2}}}\delta \left( {m{\tau_2} - t} \right). 
    \end{aligned}
\]
Therefore, functional determinant is calculated:
\begin{equation}
    \begin{aligned}
        - \log \det \left( {\slashed{\nabla} + M} \right) = \int_0^{ + \infty } {\frac{{{\mathrm{d}}t}}{t} K_{\slashed{\nabla} + M}^{\left( s \right)}\left( {\tau ;t} \right)} \\
        = \sum\limits_{m \ge 1} {\frac{{ {(-1)}^{m+1}}}{{2m{{\left| {\sin \left( {m\tau /2} \right)} \right|}^2}}}} \exp \left( { - {\rm{i}}ms{\tau _1} - mM{\tau _2}} \right).
    \end{aligned}
\end{equation}
Finally, the desired result of partition function:
\[
    \begin{aligned}
        &\log \det \left[ {\frac{{{{{\slashed\nabla }}_{\left( s \right)}} + \left( {s - 1} \right)}}{{{{{\slashed\nabla }}_{\left( {s - 1} \right)}} + s}}} \right] \\
        &= \sum\limits_{m \ge 1} {\frac{{(-1)}^{m}}{{2m{{\left| {\sin \left( {m\tau /2} \right)} \right|}^2}}}} \left[ {\exp \left( { - {\rm{i}}ms{\tau _1} - m\left( {s - 1} \right){\tau _2}} \right) - \exp \left( { - {\rm{i}}m\left( {s - 1} \right){\tau _1} - ms{\tau _2}} \right)} \right]\\
        &= \sum\limits_{m \ge 1}{(-1)}^{m} {\frac{{\exp \left( { - {\rm{i}}ms\bar \tau } \right)}}{{2m{{\left| {\sin \left( {m\tau /2} \right)} \right|}^2}}}} \exp \left( {m{\tau _2}} \right)\left[ {1 - \exp \left( {{\rm{i}}m{\tau _1} - m{\tau _2}} \right)} \right]\\
        &= \sum\limits_{m \ge 1}{(-1)}^{m} {\frac{{{{\bar q}^{ms}}}}{{m{{\left| {1 - {q^m}} \right|}^2}}}} \left[ {1 - {q^m}} \right] = 2\sum\limits_{m = 1}^\infty {(-1)}^{m} {\frac{{{{\bar q}^{ms}}}}{{m\left( {1 - {{\bar q}^m}} \right)}}}.
    \end{aligned}
\]
\begin{equation}
    \Longrightarrow {{\det }^{1/2}}\left[ {\frac{{{{{\slashed\nabla }}_{\left( s \right)}} + \left( {s - 1} \right)}}{{{{{\slashed\nabla }}_{\left( {s - 1} \right)}} + s}}} \right] = \prod\limits_{n = s}^\infty \left({1 + {{\bar q}^n}}\right).
\end{equation}
It's easy to calculate 1-loop partition function with the opposite signature of $\psi$. The result is:
\begin{equation}
    Z_{\text{HS}}^{s,\psi} = \frac{{{{\det }^{1/2}}\left( {\slashed{\nabla } - \psi (s - 1)} \right)_{\left( s \right)}^{{\rm{TT}}}}}{{{{\det }^{1/2}}\left( {\slashed{\nabla } - \psi s} \right)_{\left( {s - 1} \right)}^{{\rm{TT}}}}}
    = 
    \begin{cases}
        \prod\limits_{n = s-1/2}^\infty ({1 + {{q}^{n+1/2}}}), \psi = +,\\
        \prod\limits_{n = s-1/2}^\infty ({1 + {{\bar{q}}^{n+1/2}}}), \psi = -.
    \end{cases}
\end{equation}
Similarly,
\begin{equation}
    \begin{aligned}
        Z_{\mathrm{spinor}} &= {\det}^{1/2} {\left(\slashed{\nabla} -\zeta M \right)} \\
        &= \prod_{m,n=0} (1-q^{h+m}\bar{h}^{h-1/2+n}),\, \zeta=+, h = 1 + M/2; \\
        &\; \text{or} \prod_{m,n=0} (1-q^{h-1/2+m}\bar{h}^{h+n}),\,\zeta=-, h = 1 + M/2.
    \end{aligned}
\end{equation}
\subsection{Final result}
Combine Eq.~\eqref{eq:boson_partfunc}, Eq.~\eqref{eq:dirac_partfunc}, Eq.~\eqref{eq:boson_hs_partfunc}, Eq.~\eqref{eq:fermion_hs_partfunc} and  Eq.~\eqref{eq:majorana_chiral_partfunc1}, Eq.~\eqref{eq:majorana_chiral_partfunc2}, Eq.~\eqref{eq:fermion_chiral_hs_partfunc} together, the total partition function of our $\cn = (0, 2)$ higher-spin gravity, around thermal AdS vacuum, is given by:
\begin{equation}
    Z_{\text{1-loop}}[\nu] = \prod_{\pm_K,\pm_\psi}{(Z_{\pm_K\pm_\psi}[\nu])}^{\alpha_{\pm_K\pm_\psi}}
    \prod_{s=1}^{\infty}\prod_{i=0}^\infty \frac{{(1+{\bar{q}}^{i+s+1/2})}^2}{{|1-q^{i+s}|}^2{|1-q^{i+s+1}|}^2}.
\end{equation}where $Z_{\pm_K\pm_\psi}$ correspond to supermultiplets whose boson part lies in $K=\pm_K$, $\psi=\pm_\psi$ subspace,
\begin{equation}
    Z_{\pm_K\pm_\psi}[\nu] = Z_{\text{boson}}^{(2 + (1\mp_K \nu))/4} \cdot Z_{\text{Majorana}}^{(2 + (1\mp_K \nu))/4,\zeta= \pm_K \cdot \pm_\psi}.
\end{equation}
$\alpha_{\pm_K\pm_\psi}$ are the number of each supermultiplet contained in the higher-spin theory.

These result may serve as perturbative part of partition functions of bulk higher-spin gravity theory and guide the search for possible dual $\cn = (0, 2)$ higher-spin CFTs.

\section{Conclusion and outlook}\label{sec:discussion}
In this paper we construct a linearized higher-spin gravity theory that possibly corresponds to some $2\mathrm{d}$ $ \cn = (0, 2)$ CFTs. We discussed the representation of higher-spin algebra and find it have the structure of $2\mathrm{d}$ $ \cn = (0, 2)$
supermultiplet. Especially, we noticed the Weyl-Majorana-like behavior of fermions with the unfolded formalism of matter fields. 
We also calculated the 1-loop partition function of such a higher-spin gravity theory with different matter contents around thermal AdS vacuum by heat-kernel method.

Though the higher-spin gravity theory we studied here is only a linearized theory around some pure gravity solution, results of this paper may guide further research on $3\mathrm{d}\; \mathrm{HS}/ 2\mathrm{d}\; \mathrm{CFT}$ holography with $\cn = (0, 2)$ supersymmetry.
The completion of this linearized theory into a full interacting theory, remains to be done in further works. A natural way to do this is to construct a string theory with $\cn = (0, 2)$ worldsheet global symmetry and consider its tensionless limit. Also, a generalization of the Poisson sigma model construction raised recently~\cite{Sharapov:2024euk} may sheld light on this problem from the aspect of field theory.

Another natural next step is to consider a ``Higgs mechanism" for the higher-spin theory so that the higher-spin fields become massive and the higher-spin symmetry is broken, along the line of~\cite{Bianchi:2005ze,Gaberdiel:2015uca,Belin:2025nqd}. This transition is clearly interesting on its own right~\cite{Bekaert:2022poo},
in addition, we expect this symmetry breaking be a bulk dual of the higher-spin symmetry breaking observed in~\cite{Peng:2017kro,Peng:2018zap,Ahn:2018sgn}.\footnote{Notice that there is another integrable-chaos transition observed in SYK related models~\cite{Gao:2024lve}, and it is interesting to analyze if higher-spin symmetry and its breaking appear in that context.}
It is also interesting to investigate if there could be a Schwarzian sector in the higher-spin theory discussed in this paper, which could be dual to supersymmetric schwarzians identified in various boundary SYK-like theories~\cite{Fu:2016vas,Ghosh:2019rcj,Peng:2020euz,Datta:2021efl,Heydeman:2022lse,Benini:2024cpf,Zhang:2025kty}.

\section*{Acknowledgements}
We are grateful to Cheng Peng for introducing this research direction, as well as for his helpful discussions and consistent guidance throughout the project.
ZC is supported by NSFC NO. 12175237 and NO. 12447108, 
the Fundamental Research Funds for the Central Universities, and funds from the Chinese Academy of Sciences. 

\bibliographystyle{jhep}
\bibliography{ref}
\end{document}